\documentclass[lettersize,journal]{IEEEtran}
\usepackage{amsmath,amsfonts}
\usepackage{balance}
\usepackage{floatrow}
\usepackage{algorithm}
\usepackage{algpseudocode}
\usepackage{array}
\usepackage{subfig}
\usepackage{textcomp}
\usepackage{stfloats}
\usepackage{url}
\usepackage{verbatim}
\usepackage{graphicx}
\def\BibTeX{{\rm B\kern-.05em{\sc i\kern-.025em b}\kern-.08em
    T\kern-.1667em\lower.7ex\hbox{E}\kern-.125emX}}

\newtheorem{thm}{Theorem}
\newtheorem{defn}{Definition}

\begin{document}
\title{High-Density Coding Scheme for SWIPT Systems}
\author{Dongheon~Lee, Gyuyeol~Kong, Jang-Won~Lee, and Sooyong~Choi}%


\maketitle

\begin{abstract}
In this study, a novel coding scheme called high-density coding based on high-density codebooks using a genetic local search algorithm is proposed.
The high-density codebook maximizes the energy transfer capability by maximizing the ratio of 1 in the codebook while satisfying the conditions of a codeword with length $n$, a codebook with $2^k$ codewords, and a minimum Hamming distance of the codebook of $d$.
Furthermore, the proposed high-density codebook provides a trade-off between the throughput and harvested energy with respect to $n$, $k$, and $d$.
The block error rate performances of the designed high-density codebooks are derived theoretically and compared with the simulation results.
The simulation results indicate that as $d$ and $k$ decrease, the throughput decreases by a maximum of $10\%$ and $40\%$, whereas the harvested energy per time increases by a maximum of $40\%$ and $100\%$, respectively.
When $n$ increases, the throughput decreases by a maximum of $30\%$, while the harvested energy per time increases by a maximum of $110\%$.
With the proposed high-density coding scheme, the throughput and harvested energy at the user can be controlled adaptively according to the system requirements.
\end{abstract}


\section{Introduction}
\IEEEPARstart{I}{n} the era of Internet of Things (IoT), the role of wireless devices is emerging \cite{ref:IoT}.
To prolong the lifetime of the IoT network, batteries of wireless devices must be replaced or recharged.
However, the management of batteries for a large number of wireless devices is expensive and inconvenient \cite{ref:precoder}.
Therefore, simultaneous wireless information and power transfer (SWIPT) systems have received an extensive amount of interest to achieve the efficient power management of batteries.
Studies on SWIPT systems have been conducted for various applications, such as the design of a waveform \cite{ref:waveform} or precoder \cite{ref:precoder, ref:SCLNR}, and channel coding \cite{ref:magazine, ref:unary, ref:RLL1, ref:RLL2} \cite{ref:SWIPT}.

In particular, the coding scheme for SWIPT systems differ from conventional coding schemes for wireless information transfer (WIT) systems.
Conventional channel coding schemes, such as Hamming codes, which have been widely used for WIT systems, do not consider the energy-harvesting performance.
Instead, the Hamming code is designed to achieve the highest possible rate with a minimum Hamming distance of 3 between codewords in a codebook to increase the error detection and correction capabilities\cite{ref:minimum_distance}.
In wireless power transfer systems wherein the signal transmission time and power are critical, energy-harvesting performance degradation occurs when the conventional coding method is used.
Therefore, to improve the energy-harvesting performance of the channel coding scheme in SWIPT systems, several studies have been conducted \cite{ref:magazine, ref:unary, ref:RLL1, ref:RLL2}.

The impact of coding on SWIPT systems was discussed in \cite{ref:magazine}, which introduced a design principle for SWIPT coding.
In \cite{ref:unary}, a unary-coded SWIPT transceiver was analyzed from an information theoretical perspective.
In \cite{ref:RLL1}, a run-length-limited (RLL) encoder was designed to minimize the battery overflow/underflow probability at the receiver subject to a constraint on the achievable mutual information.
Moreover, the capacity lower bound of the BSC and Z-channel was determined in \cite{ref:RLL2} when an  RLL code was adopted.

Most of the research on coding schemes for SWIPT systems is based on systems using on--off keying (OOK) modulation.
The OOK modulation uses two types of signals, namely `on' and `off,' because analog modulation schemes are considered to be efficient for SWIPT systems \cite{ref:modulation}.
Analog modulation schemes do not require a local oscillator or an analog to digital converter, which consume relatively high power; moreover, OOK modulation is the simplest modulation among analog modulations \cite{ref:modulation}.
The ratio of the `on' signal to the `off' signal is critical for energy harvesting because energy harvesters can only harvest energy from the `on' signal.
Therefore, maintaining a ratio of 1 is vital in codeword design.
To increase the number of 1's in a codeword, the unary code and RLL code were developed in \cite{ref:unary, ref:RLL1}; these codes are a type of variable length code requiring complex decoding algorithms \cite{ref:VLC}.

In this study, we propose a high-density channel coding scheme and a high-density codebook design algorithm for SWIPT systems using OOK modulation.
The proposed high-density codebook design algorithm produces a fixed-length codebook that can decode with a lower complexity than that in the case of an RLL code or unary code.
In addition, the proposed high-density codebook satisfies the following two conditions: maximization of the ratio of 1 in the codebook and preservation of the minimum Hamming distance between the codewords in the codebook using a genetic search algorithm.

The proposed high-density codebook can be obtained using a genetic local search algorithm that comprises three steps, namely local search, recombination, and final selection.
In the local search step, a genetic local search algorithm produces codewords that satisfy the given condition, such as the length of the codeword, number of codewords in the codebook, and minimum Hamming distance in the codebook.
The recombination step generates new codebooks by transforming the codebooks from the local search step.
The final selection step selects the codebook with the maximum number of 1's.
Therefore, the proposed coding scheme can increase the duration of actual transmit RF signals and maintain the throughput of information transmission while achieving
performance gain in terms of harvested energy.

In addition, the proposed high-density codebooks demonstrate a trade-off between the throughput and harvested energy performance based on various codebook design parameters such as codeword length, number of codewords in the codebook, and minimum Hamming distance of the codebook.
Therefore, it is possible to transmit a signal adaptive to conditions such as the data and energy requirements of the receiver and the bit error rate constraint of the systems.

%

\section{System Model}
\label{section:systemmodel}


\begin{figure}[!t]
\centering
\includegraphics[width=0.9\columnwidth] {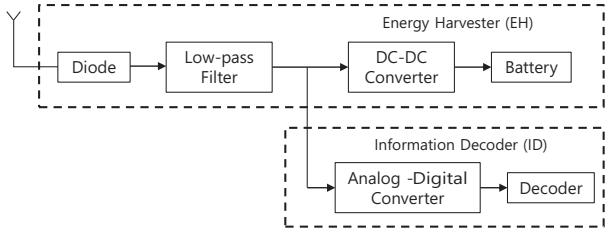}
\caption{Receiver for both information decoding and energy harvesting.}
\label{fig:receiver}
\end{figure}
We consider single-input single-output SWIPT systems using OOK modulation.
An access point (AP) transmits the OOK signal, and both the AP and receiver comprise the codebook information and block error rate (BLER) table of each codebook based on the signal-to-noise ratio (SNR).
An AP transmits the OOK signals used at the receiver for information decoding and energy harvesting, as illustrated in Fig. \ref{fig:receiver}.
Fig. \ref{fig:receiver} depicts the receiver structure considered in this study.
The received signal at the antenna is converted into a DC signal using a rectifier.
Because the RF to DC conversion in the rectifier is analogous to the RF band to baseband conversion, a mixer that consumes high energy is not required for information decoding \cite{ref:receiver}.

To analyze the relationship between the BLER and harvested energy among the codebooks in the additive white Gaussian noise channel, we consider the received signal at the receiver as
\begin{align}
y = x + z, \nonumber
\end{align}
where $x \in \left\{ {0,\sqrt {2 \cdot E_b } } \right\}$ denotes the OOK signal with a bit energy $E_b$ and $z \sim {\cal N}\left( {0,N_0 /2} \right)$ denotes the identically distributed additive Gaussian noise.
In addition, it is assumed that the maximum likelihood (ML) decoding for soft decisions is used for the codeword estimator.
Then, the codebook can be expressed as ${(n,k,d)}$ with $W$, where $n$ denotes the length of the codeword in the codebook, $2^k$ denotes the number of codewords in the codebook, $d$ denotes the minimum Hamming distance between two different codewords in the codebook, and $W$ denotes the number of 1's in the codebook.
The performance metrics of the codebook can be both throughput and harvested energy.
The throughput for the ${(n,k,d)}$ codebook system can be defined as
\begin{equation}
{\eta _{n,k,d}} = \frac{k}{n} \times \left( {1 - {{P}_{n,k,d}}} \right).
\end{equation}
where ${{P}_{n,k,d}}$ denotes the BLER of the ${(n,k,d)}$ codebook.
The theoretical BLER of the designed ${(n,k,d)}$ codebook from the proposed high-density codebook design algorithm can be given by
\begin{equation}
\begin{array}{l}
P_{n,k,d}  \approx \sum\limits_{i = 0}^{2^k  - 1} {\frac{1}{{2^k }} \cdot A_{d,i}  \cdot Q\left( {\sqrt {d \cdot \frac{{E_b }}{{N_0 }}} } \right)}.
\end{array}
\label{ref:BLER}
\end{equation}
where $A_{d,i}$ denotes the number of codewords that have the Hamming distance $d$ with the $i$th codeword in the codebook.
The proof is presented in the Appendix.

Additionally, the harvested energy for the ${(n,k,d)}$ codebook system can be measured in two ways:
\begin{equation}
Q_{n,k,d}^b  = \frac{{n_{n,k,d} }}{k}.
\end{equation}
\begin{equation}
Q_{n,k,d}^t  = \frac{{n_{n,k,d} }}{n}.
\end{equation}
where ${Q_{n,k,d}^b}$ denotes the harvested energy per bit, ${Q_{n,k,d}^t}$ denotes the harvested energy per time, and ${n_{n,k,d}}$ denotes the number of 1's in the ${(n,k,d)}$ codebook.
In Section \ref{section:systemmodel}, we evaluate several codebooks generated from the proposed algorithm to compare both the BLER and harvested energy performances presented in Section \ref{section:simulaionresults}.
\\

\section{Proposed Channel Coding Scheme and Codebook Design Algorithm}
\label{section:HDC}

\begin{figure}[!t]
\centering
\subfloat[]{
\includegraphics[width=0.4\columnwidth] {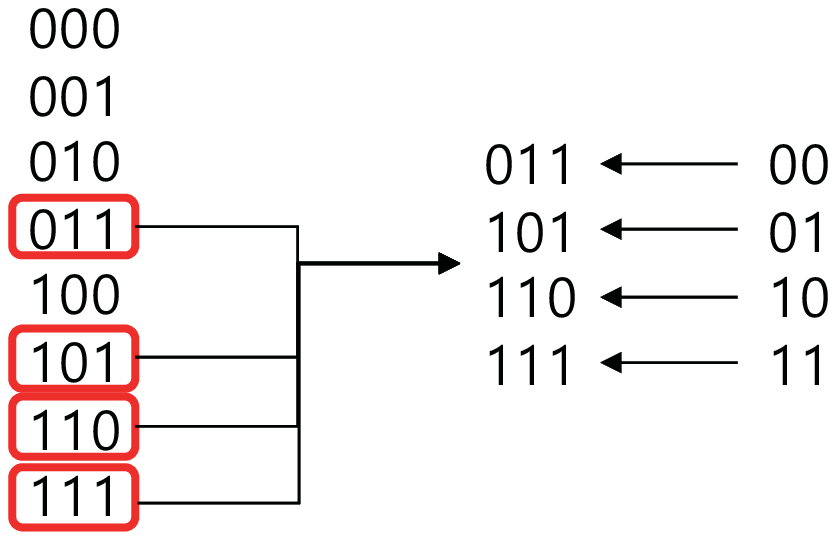}
\label{subfig:trade-off1}\
}
\subfloat[]{
\includegraphics[width=0.4\columnwidth] {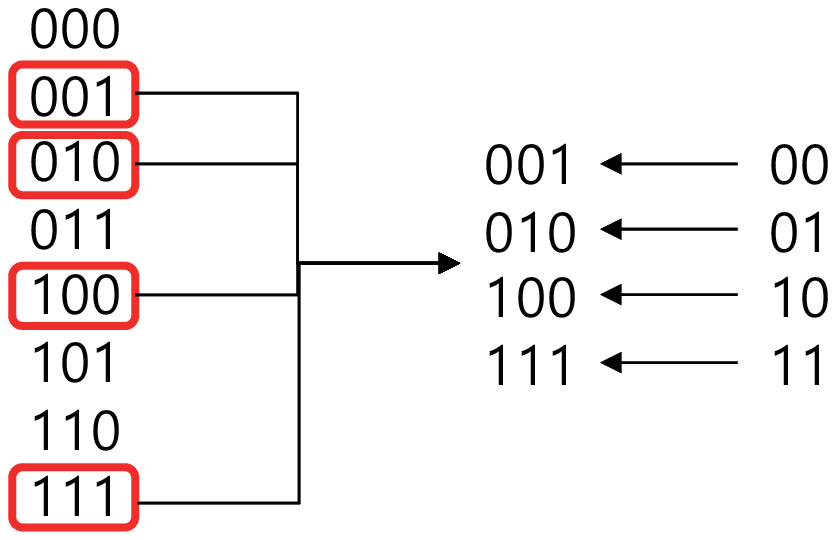}
\label{subfig:trade-off2}\
}
\caption{
Codeword selection methods for $(3,2,d)$ codebook (a) $W = 9$, $d = 1$ and (b) $W = 6$, $d = 2$.
}
\label{fig:trade-off}
\end{figure}

\subsection{High-Density Coding Scheme}
High-density coding encodes $k$ bits into $n$ bits of as many 1's as possible.
The encoded bits are transmitted in the communication system using OOK modulation, where a digital bit 1 indicates the presence of a carrier wave and digital bit 0 indicates the absence of a carrier wave.
Therefore, the number of 1's represents the number of transmissions and determines the energy-harvesting performance.
In addition, the error probability depends on the minimum distance of the codewords, which is related to the Hamming weights \cite{ref:minimum_distance}.
Therefore, the number of 1's in a codebook can determine the energy-harvesting performance as well as error probability performance. 

Fig. \ref{fig:trade-off} illustrates an example of the trade-off between the number of 1's and the minimum Hamming distance.
In Fig. \ref{fig:trade-off}, the codewords with the largest number of 1's satisfying the condition of minimum Hamming distance $d$ are selected to construct the $(3,2,d)$ codebook.
The energy-harvesting performance of the codebook depicted in Fig. \ref{subfig:trade-off1} is better than that of the codebook depicted in Fig. \ref{subfig:trade-off2} because the codebook depicted in Fig. \ref{subfig:trade-off1} has a larger number of 1's than that of the codebook depicted in Fig. \ref{subfig:trade-off2}.
However, the BLER performance of the codebook depicted in Fig. \ref{subfig:trade-off2} is better than that of the codebook depicted in Fig. \ref{subfig:trade-off1} because the codebook depicted in Fig. \ref{subfig:trade-off2} has a larger minimum distance than that of the codebook depicted in Fig. \ref{subfig:trade-off1}.

Therefore, we propose a high-density coding scheme that uses a codebook with the maximum number of 1's for maximizing the harvested energy while satisfying the minimum Hamming distance of the codebook for maintaining error correction capability.
Thus, unlike the conventional communication systems that select a codebook solely based on the consideration of an error correction capability or a BLER, the proposed channel coding scheme selects a codebook considering the harvested energy.

However, in general, designing a codebook for a given ${(n,k,d)}$ is an NP--hard problem that is challenging, particularly for large values of $n$, $k$, and $d$ \cite{ref:NP-hard}.
In addition to the ${(n,k,d)}$ condition, we consider the performance gain of the harvested energy by maximizing the number of 1's in the codebook.
Therefore, designing a codebook that satisfies all the conditions is even more challenging.
To this end, a high-density codebook design algorithm using a genetic local search algorithm was proposed.
The proposed codebook design algorithm using the genetic local search algorithm is described in Section~\ref{subsection:GLSA}.

\subsection{Genetic Local Search-Based High-Density Codebook Design}
\label{subsection:GLSA}

\begin{figure}[!t]
\centering
\includegraphics[width=0.9\columnwidth] {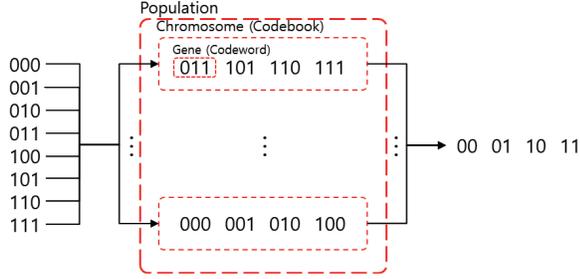}
\caption{
Parameter mapping between coding scheme and genetic algorithm for $(3,2,1)$ codebook.
}
\label{fig:GA}
\end{figure}

In this section, we introduce the proposed channel coding scheme and codebook design algorithm for SWIPT systems.
The codebooks are designed to maximize the number of 1's in the codebook while achieving the minimum distance between the codewords.
The codebook design algorithm based on the genetic local search algorithm constructs a codebook with the maximum number of 1's while maintaining the minimum distance between the codewords.
Fig. \ref{fig:GA} depicts a relationship between the parameters of the coding scheme and genetic algorithm.
The population of the genetic local search algorithm comprises a set of codebooks.
The codebook and codeword correspond to the chromosome and gene in a conventional genetic algorithm, respectively.

The genetic local search algorithm begins by generating an initial population ${P_{input}}$ and then modifies the population at each step.
First, the population ${P}$ increases by applying the local search procedure to each codebook ${C}$ in ${P}$.
Next, the enlarged population ${P'}$ is generated by producing child codebooks through the application of a recombination procedure to the population ${P}$.
The population ${P'}$ increases by applying the local search procedure to each codebook ${C}$ in ${P'}$, and a certain number of best codebooks are selected.
The RECOMBINATION, second LOCAL SEARCH, and SELECTION algorithms are repeated until the stop criterion is satisfied.
The entire algorithm is summarized in Algorithm 1.
In the following subsections, we describe the details of the local genetic search algorithm.

\begin{algorithm}
    \caption{GENETIC LOCAL SEARCH}
    \label{algorithm:GENETIC LOCAL SEARCH}
    \begin{algorithmic}
        \Require{Initial population ${P_{input}}$}
            \State {$P = {P_{input}}$}
            \ForAll {$C \in P$}
            \State {$C =$ LOCAL SEARCH$\left( C \right)$}
            \EndFor
            \Repeat
            \State {$P' =$ RECOMBINATION$\left( P \right)$}
            \ForAll {$C \in P'$}
            \State {$C =$ LOCAL SEARCH$\left( C \right)$}
            \State {$P = P \cup \left\{ C \right\}$}
            \EndFor
            \State {$P =$ SELECTION$\left( P \right)$}
            \Until {STOP}
    \end{algorithmic}
\end{algorithm}

%

\subsubsection{Initial population}
The initial population ${P_{input}}$ in the genetic local search algorithm is randomly generated.
The initial population comprises a fixed number of codebooks.
Each codebook consists of a number of codewords between 1 and $2^k  - 1$, satisfying the minimum Hamming distance $d$.
Owing to the complexity of generating the initial population, it is preferable not to generate an excessively large number of codewords in the initial population.
On the contrary, if there are extremely few codewords in the initial population, sufficient population diversity cannot be achieved.

\subsubsection{Local search}
In this step, we describe a local search procedure in Algorithm \ref{algorithm:GENETIC LOCAL SEARCH}, which extends the input codebooks to obtain a large number of codewords.
The codebook, which includes a large number of codewords, exhibit a high probability of the selected ${k}$ codewords having a large number of 1's.
Because the local search procedure is operated by satisfying the minimum distance between the codebooks, the codebook design condition for minimum distance can be maintained.

The inputs for the LOCAL SEARCH algorithm are the codebook ${C_{input}}$ and the position ${S}$, which denotes the position indicator in the codeword.
First, the mutation procedure is applied to the codebook ${C}$ at position ${S}$.
Next, when the input codebook is $C$, from ${\mathbf{x}} = {\mathbf{0}}$ to ${\mathbf{x}} = {\mathbf{1}}$, $\mathrm{d}\left( {{\mathbf{x}},C} \right)$, which denotes the Hamming distance between ${\mathbf{x}}$ and $C$, in each iteration can be calculated.
When $\mathrm{d}\left( {{\mathbf{x}},C} \right) \ge d$, $C$ is extended by including ${\mathbf{x}}$.
When $\mathrm{d}\left( {{\mathbf{x}},C} \right) \le d$ , ${\mathbf{x}}$ becomes ${\mathbf{x}} + 1$.
The extending procedure is repeated until ${\mathbf{x}} = {\mathbf{1}}$.

\begin {defn}
Lexicographical ordering, $\le_L$, is defined as for any ${\mathbf{x}},{\mathbf{y}} \in \mathbb{Z}^n$,
\begin{align}
{\mathbf{x}} \le _L {\mathbf{y}} &\Leftrightarrow ({\mathbf{x}} = {\mathbf{y}}) \nonumber \\
& \vee (\exists 1 \le l \le n:x_i  = y_i ,i = 1,2, \ldots ,l - 1,x_l  < y_l ). \nonumber
\end{align}

\end {defn}

\noindent If $C$ is extended in this manner, any ${\mathbf{y}}$ with ${\mathbf{x}}{ \le _L}{\mathbf{y}}$ exhibits a smaller Hamming distance to $C$ than $d$.
Therefore, extending the codebook suffices to continue the procedure with ${\mathbf{x}} + 1$.
For a given input codebook $C$, the codebook-extending procedure always yields the same extended codebook.
Therefore, for a variety of populations, a process that generates a new codebook $C'$ transformed from codebook $C$, the so-called mutation, is required.
In the mutation process, the positions in the codebook with a length of $n$ are randomly selected, and the bits in the selected positions for every codeword in the codebook $C$ are modified by swapping the values 1 and 0 with each other.
Because the mutation does not affect the $\left( {n,k,d} \right)$ property of the codebook $C$, the codebook $C$ is equivalent to the transformed codebook $C'$.
After applying the local search algorithm to $C'$, the inverse of the mutation applied to codebook $C$ is applied to the output of the codebook extension procedure.
Applying the inverse of the mutation process causes the extended codebook to include the input codebook $C$.
The local search algorithm is summarized below.

\begin{algorithm}
    \caption{LOCAL SEARCH}
    \label{algorithm:LOCAL SEARCH}
    \begin{algorithmic}
        \Require{Codebook ${C_{input}}$, Positions $S$}
            \State {$C = {C_{input}}$}
            \State {$C' = mutation\left( {C,S} \right) \triangleleft$ bit flip at the positions $S$ in $C$}
            \State {${\mathbf{x}} = {\mathbf{0}}$}
            \Repeat
            \If {$\mathrm{d}\left( {{\mathbf{x}},C'} \right) \ge d$}
            \State {$C' = C' \cup \left\{ {\mathbf{x}} \right\}$}
            \EndIf
            \State {${\mathbf{x}} = {\mathbf{x}} + 1$}
            \Until {${\mathbf{x}} = {\mathbf{1}}$}
            \State {$C = mutation\left( {C',S} \right)$}
    \end{algorithmic}
\end{algorithm}

\subsubsection{Recombination}
In this step, a new codebook is generated from the output codebook of the LOCAL SEARCH algorithm.
Recombination is the most significant process for a local genetic search algorithm for achieving a sub--optimal solution.
Recombination extends the search area and therefore increases the diversity of the population.

We assume that all input populations for RECOMBINATION have the same $p$ even number of codewords.
We first select $p/2$ pairs of parent codebooks from the input population ${P_{input}}$, such that codebooks with a large number of 1's are selected as a parent codebook with a higher probability.
If the number of codewords $m$ in the ${i^{th}}$ codebook is smaller than $2^k$, the number of 1's in the ${i^{th}}$ codebook multiplied by $2^k/m$ represents the weight ${W_i}$ of the codebook.
Otherwise, the number of 1's represents the weight ${W_i}$ of the codebook.
If ${W_{\min }}$ is the minimum weight among ${W_i}$, $i \in \left\{ {1,...,p} \right\}$, the ${i^{th}}$ codebook is selected as a parent with probability $p_i$ written as
\begin{equation}
\begin{array}{l}
p_i  = 1/c \times \left( {W_i  - W_{\min }  + 1} \right).
\end{array}
\label{ref:probability}
\end{equation}
where $c = \sum\nolimits_{j = 1}^p {\left( {{W_j} - {W_{\min }} + 1} \right)} $.
Next, we combine the two codebooks to produce the child codebooks.

\begin{thm}
\emph{Let ${C_j}$ and ${C_k}$ be two codebooks with length $n$ and a minimum Hamming distance of at least $d$.
Let ${\mathbf{a}}$ be a codeword in $\mathbb{Z}^n$, and $D$ be an integer with $0 \le D \le n + d$. Then,
\begin{equation}
\begin{array}{l}
\left\{ {{\mathbf{x}} \in {C_j}|\mathrm{d}\left( {{\mathbf{x}},{\mathbf{a}}} \right) \le D - d} \right\} \cup \left\{ {{\mathbf{x}} \in {C_k}|\mathrm{d}\left( {{\mathbf{x}},{\mathbf{a}}} \right) \ge D} \right\} \nonumber
\end{array}
\end{equation}
is a codebook with length $n$ and a minimum Hamming distance of at least $d$.}
\end{thm}

\begin{IEEEproof}
Let ${\mathbf{x_1}}$ be a codeword from the codebook $\left\{ {{\mathbf{x}} \in {C_j}|\mathrm{d}\left({{\mathbf{x}},{\mathbf{a}}} \right) \le D-d} \right\}$ and ${\mathbf{x_2}}$ be a codeword from the codebook $\left\{ {{\mathbf{x}} \in {C_k}|\mathrm{d}\left({{\mathbf{x}},{\mathbf{a}}} \right) \ge D} \right\}$.
To ensure that $\mathrm{d}\left({{\mathbf{x_1}},{\mathbf{x_2}}} \right)$ is smaller than $d$, $\left| {\mathrm{d}\left( {{\mathbf{x}}_1 ,{\mathbf{a}}} \right) - \mathrm{d}\left( {{\mathbf{x}}_2 ,{\mathbf{a}}} \right)} \right|$ should be smaller than $d$.
However, the minimum value of $\left| {\mathrm{d}\left( {{\bf{x}}_1 ,{\mathbf{a}}} \right) - \mathrm{d}\left( {{\bf{x}}_2 ,{\mathbf{a}}} \right)} \right|$ is $d$ when $\mathrm{d}\left( {{\mathbf{x}}_1 ,{\mathbf{a}}} \right)$ has the largest value $D-d$, and $\mathrm{d}\left( {{\mathbf{x}}_2 ,{\mathbf{a}}} \right)$ has the smallest value of $D$.
\end{IEEEproof}

\noindent According to the above theorem, the produced child codebooks have the same length and minimum Hamming distance as those of the parent codebooks.
Let ${C_1}$ and ${C_2}$ be a selected parent pair of codebooks; select the codeword ${\mathbf{a}}$ and integer $D$ that satisfy the conditions of length $n$ and $0 \le D \le n + d$.
According to ${C_1}$, ${C_2}$, ${\mathbf{a}}$, and $D$, the two child codebooks $C_1 '$ and $C_2 '$ are generated as follows:
\begin{align}
C_1 '&= \left\{ {{\mathbf{x}} \in {C_1}|\mathrm{d}\left( {{\mathbf{x}},{\mathbf{a}}} \right) \le D - d} \right\} \cup \left\{ {{\mathbf{x}} \in {C_2}|\mathrm{d}\left( {{\mathbf{x}},{\mathbf{a}}} \right) \ge D} \right\} \nonumber \\
C_2 '&= \left\{ {{\mathbf{x}} \in {C_2}|\mathrm{d}\left( {{\mathbf{x}},{\mathbf{a}}} \right) \le D - d} \right\} \cup \left\{ {{\mathbf{x}} \in {C_1}|\mathrm{d}\left( {{\mathbf{x}},{\mathbf{a}}} \right) \ge D} \right\},
\label{ref:recombination}
\end{align}
Then, we include $C_1 '$ and $C_2 '$ in a new population $p'$.
The selection of parent codebooks and generation of child codebooks are repeated until the $p/2$ pair of child codebooks is produced.
Therefore, $p$ child codebooks can be generated from $p$ parent codebooks.
The recombination algorithm is summarized below.
\begin{algorithm}
    \caption{RECOMBINATION}
    \label{c4.5}
    \begin{algorithmic}
        \Require{Population ${P_{input}}$}
            \State {$P = {P_{input}}$}
            \State {$P' = \left\{ {} \right\}$}
            \State {$i = 1$}
            \Repeat
            \State {select {$C_1 ,C_2  \in P$} with the probability in equation (\ref{ref:probability})}
            \State {produce $C_1 '$ and $C_2 '$ using equation (\ref{ref:recombination})}
            \State {$P' = P' \cup \left\{ {C_1 ',C_2 '} \right\}$}
            \State {$i = i + 1$}
            \Until {$i = p/2$}
    \end{algorithmic}
\end{algorithm}

\subsubsection{Selection}
In the SELECTION algorithm, $p$ codebooks are selected, which can serve as the input codebooks for the subsequent iteration of the algorithm, from the $p$ parent codebooks in population $P$ and $p$ child codebooks in population $P'$.
If there is no codebook with more than $2^k$ codewords, we select the $p$ codebooks that have the largest ${W_i}$.
If there are more than $p/2$ codebooks with more than $2^k$ codewords, we select $p/2$ codebooks that have the largest ${W_i}$ among the codebooks that have more than $2^k$ codewords, and select the remaining $p/2$ codebooks that have the largest ${W_i}$ among the codebooks that have less than $2^k$ codewords.
If there are $q\left( { \le p/2} \right)$ codebooks with more than $2^k$ codewords, we select $q$ codebooks with more than $2^k$ codewords and select the remaining $p-q$ codebooks that have the largest ${W_i}$ among the codebooks with less than $2^k$ codewords.

\subsubsection{Stop criterion}
After the selection step, the stop criterion decides whether to repeat or stop the local genetic search algorithm.
If there are codebooks with more than $2^k$ codewords, we stop the algorithm until the maximum codebook weight among the codebooks in a population does not change in a certain number of iterations.
If there is no codebook with more than $2^k$ codewords, we stop the algorithm until the maximum number of codewords among the codebooks in a population does not change in a certain number of iterations.

\section{Simulation Results}
\label{section:simulaionresults}

\begin{figure}[!t]
\centering
\hspace*{-60pt}
\subfloat[]{
\includegraphics[width=0.55\linewidth] {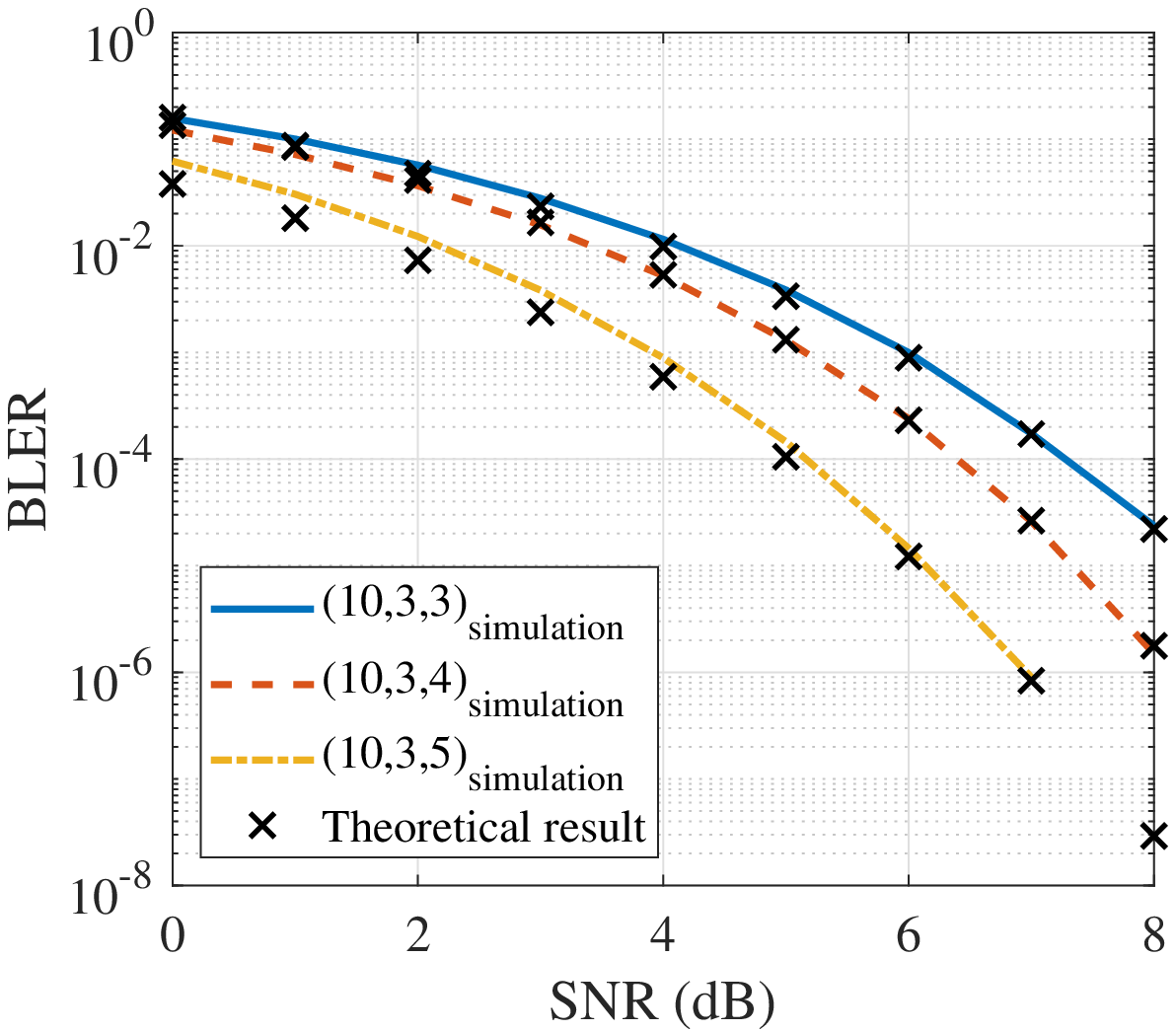}
\label{subfig:BLER_d}\
}
\hspace*{-20pt}
\subfloat[]{
\includegraphics[width=0.55\linewidth] {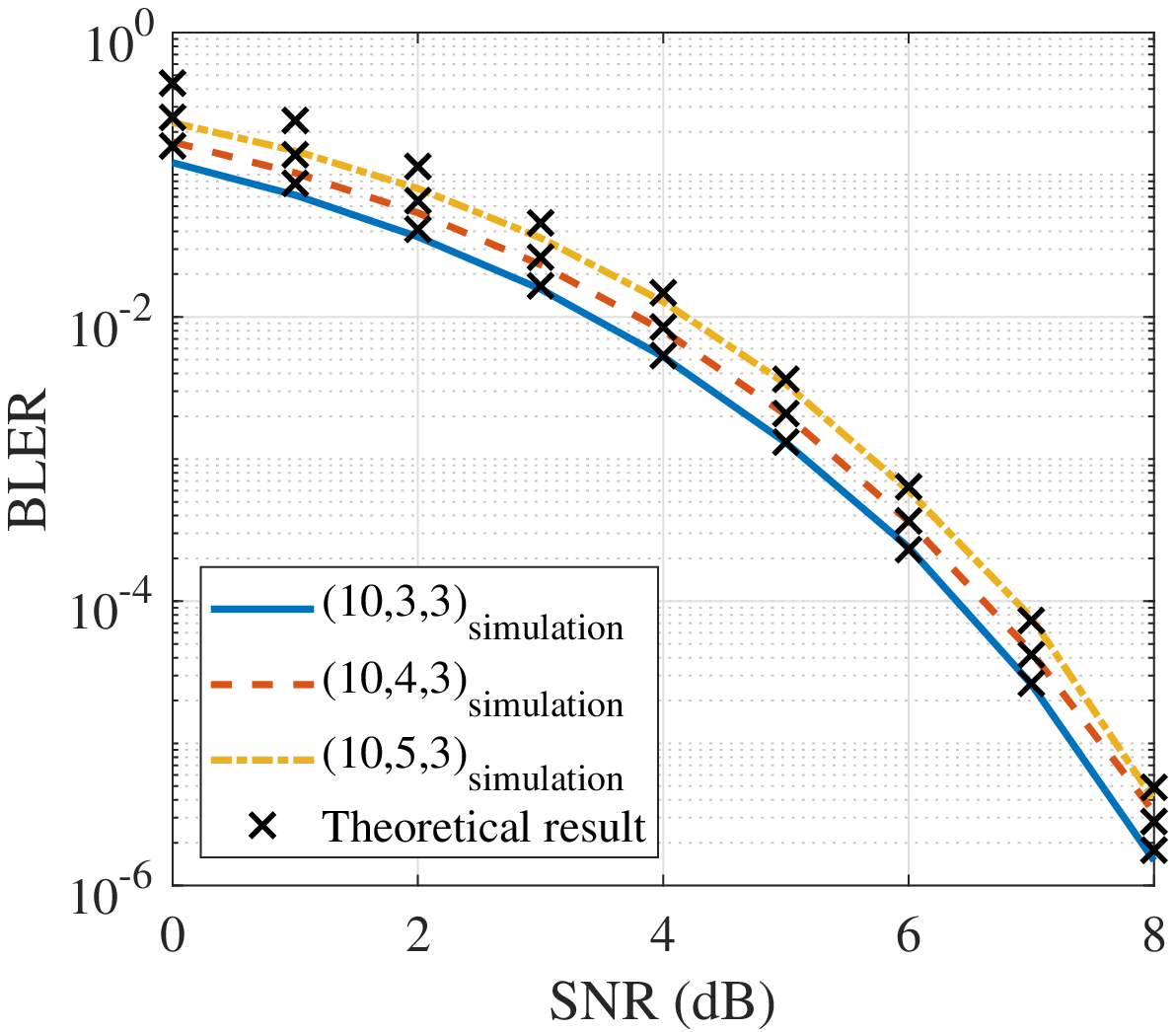}
\label{subfig:BLER_k}\
}
\hspace*{-70pt}
\caption{
BLER performance with varying SNR. (a) Difference in the minimum Hamming distance, $d$. (b) Difference in the number of codewords, $2^k$.
}
\label{fig:BLER}
\end{figure}

\begin{figure}[!t]
\centering
\hspace*{-60pt}
\subfloat[]{
\includegraphics[width=0.55\linewidth] {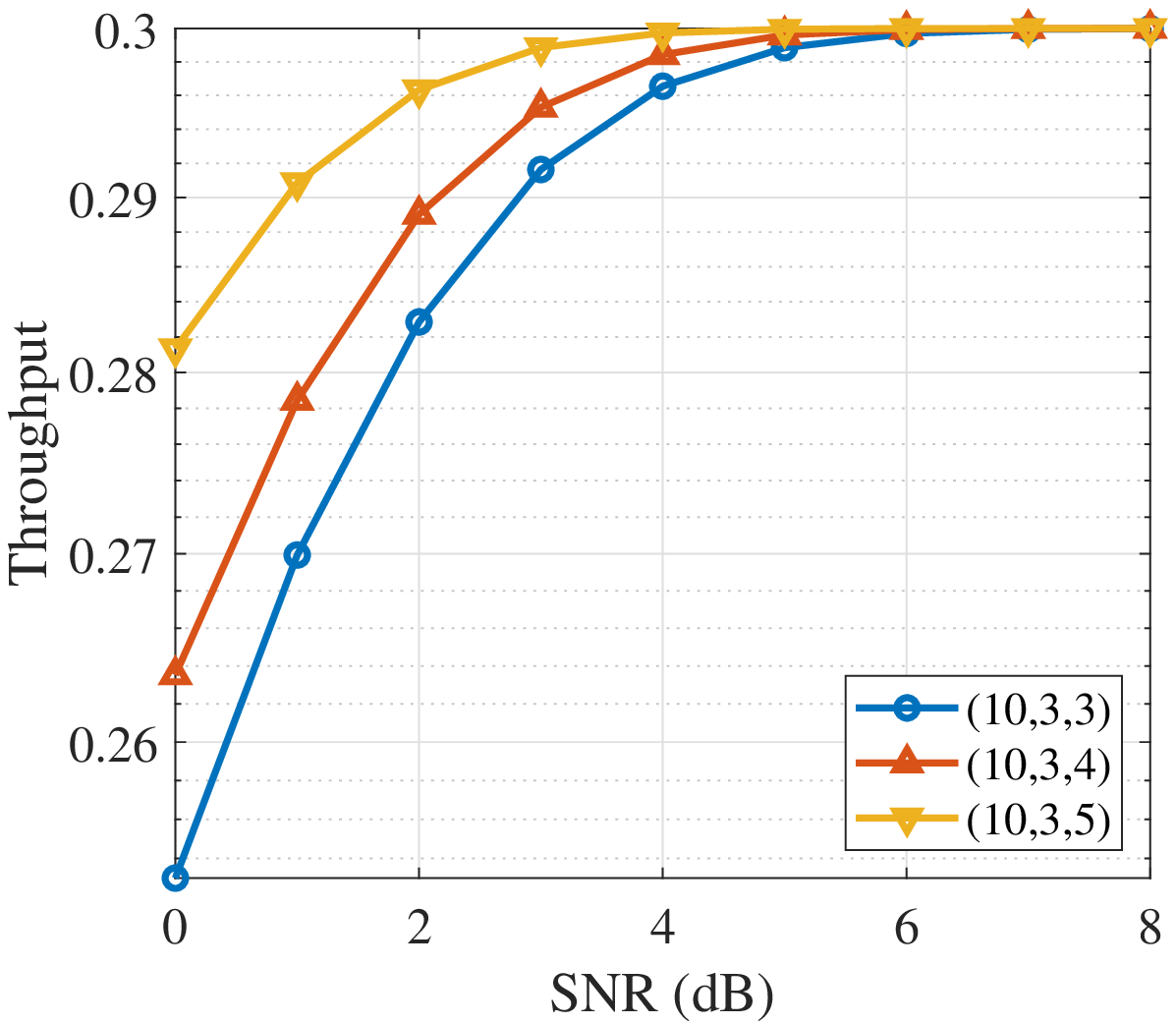}
\label{subfig:Tradeoff_d_throughput}\
}
\hspace*{-20pt}
\subfloat[]{
\includegraphics[width=0.55\linewidth] {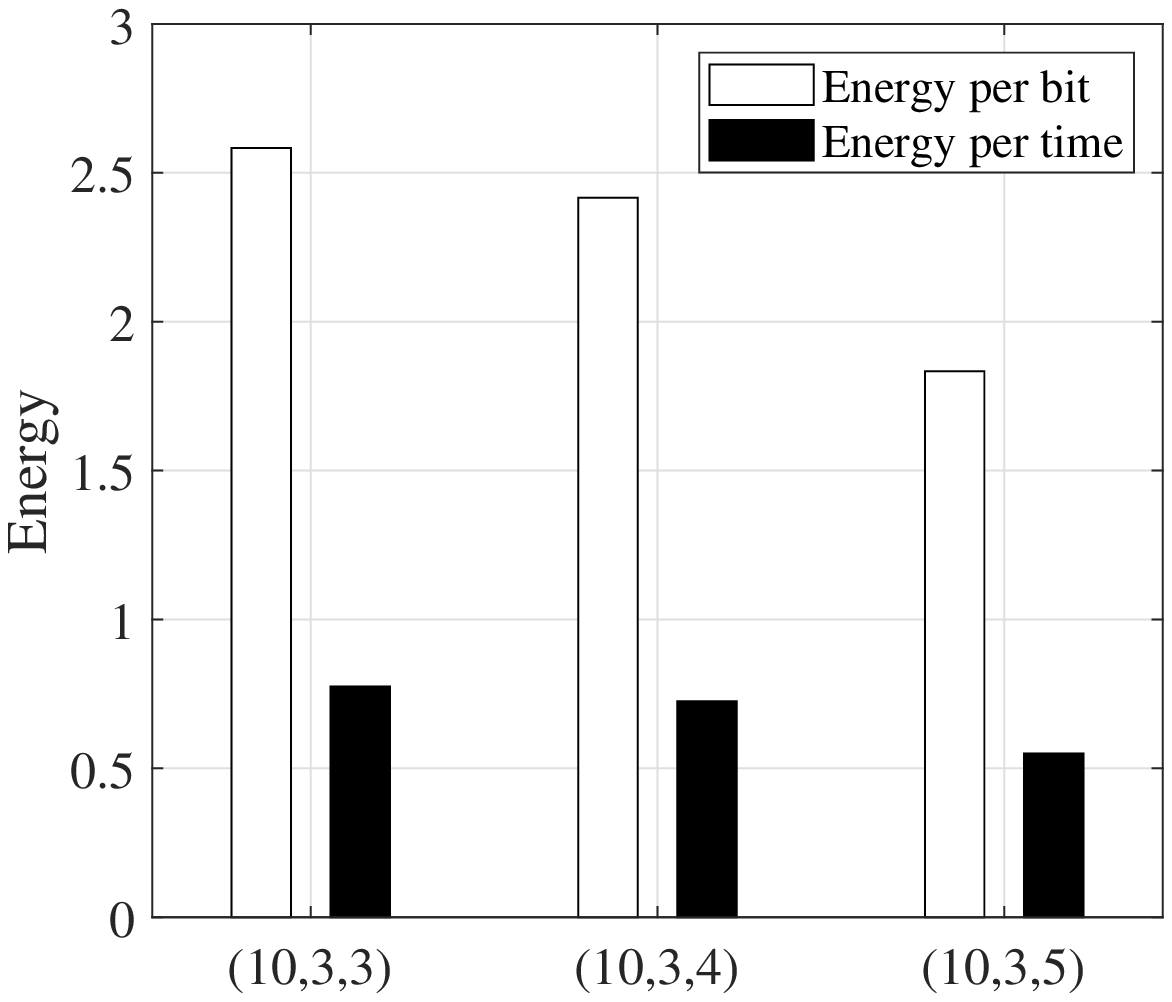}
\label{subfig:Tradeoff_d_energy}\
}
\hspace*{-70pt}
\caption{
Performance trade-off with the difference in the minimum Hamming distance, $d$. (a) Throughput ${\eta _{10,3,d}}$. (b) Energy per time, $Q_{10,3,d}^t$, and energy per bit, $Q_{10,3,d}^d$.
}
\label{fig:Tradeoff_d}
\end{figure}

In this section, we evaluate the proposed high-density coding in SWIPT systems.
The $\left( {n,k,d} \right)$ codebooks using the proposed high-density codebook design algorithm are evaluated in terms of the BLER for the theoretical results.
The number of codebooks for the initial population and number of selected codebooks in each iteration $p$ are 10.
The number of codewords in each codebook is randomly determined between 1 and 5.
The SNR at the receiver varies from 0 to 8 dB.
All the results are averaged over $10^7$ channel realizations.

In Fig. \ref{subfig:BLER_d}, the $(10,3,3)$, $(10,3,4)$, and $(10,3,5)$ codebooks are evaluated in terms of the BLERs for the simulation and theoretical results according to various SNRs for different minimum Hamming distances $d$ with fixed values of $n$ and $k$.
Fig. \ref{subfig:BLER_d} depicts that the BLER decreases as the minimum Hamming distance $d$ increases.
As the minimum Hamming distance in the codebook increases, the difference between the likelihoods of the codewords in ML decoding increases.
Based on the theoretical results in (\ref{ref:BLER}), as the minimum Hamming distance in the codebook increases, the value of the Q function decreases.
In Fig. \ref{subfig:BLER_k}, the $(10,3,3)$, $(10,4,3)$, and $(10,5,3)$ codebooks are evaluated in terms of the BLERs for the simulation and theoretical results according to various SNRs for different numbers of codewords $k$ with fixed values of $n$ and $d$.
Fig. \ref{subfig:BLER_k} depicts that the BLER increases as the number of codewords $k$ increases.
As the number of codeword pairs with a minimum Hamming distance increases, the number of detection errors in ML decoding increases.
Based on the theoretical results in (\ref{ref:BLER}), as the number of codewords in the codebook increases, the probability of increasing the value of ${A_{d,i}}$ increases.

In Fig. \ref{subfig:Tradeoff_d_throughput} and Fig. \ref{subfig:Tradeoff_d_energy}, the $(10,3,3)$, $(10,3,4)$, and $(10,3,5)$ codebooks are evaluated in terms of throughput and harvested energy for different minimum Hamming distances $d$ with fixed values of $n$ and $k$, respectively.
The $(10,3,3)$ codebook yields the highest performance in terms of harvested energy and lowest performance in terms of throughput.
In contrast, the $(10,3,5)$ codebook yields the lowest performance in terms of harvested energy and highest performance in terms of throughput.
As $d$ increases, the throughput increases because the BLER decreases while the harvested energy increases; this is because the number of 1's in the proposed high-density codebook increases.
The harvested energy per time increases by a maximum of $40\%$, and the harvested energy per bit increases by a maximum of $35\%$, while the throughput decreases by a maximum of $10\%$.

\begin{figure}[!t]
\centering
\hspace*{-60pt}
\subfloat[]{
\includegraphics[width=0.55\linewidth] {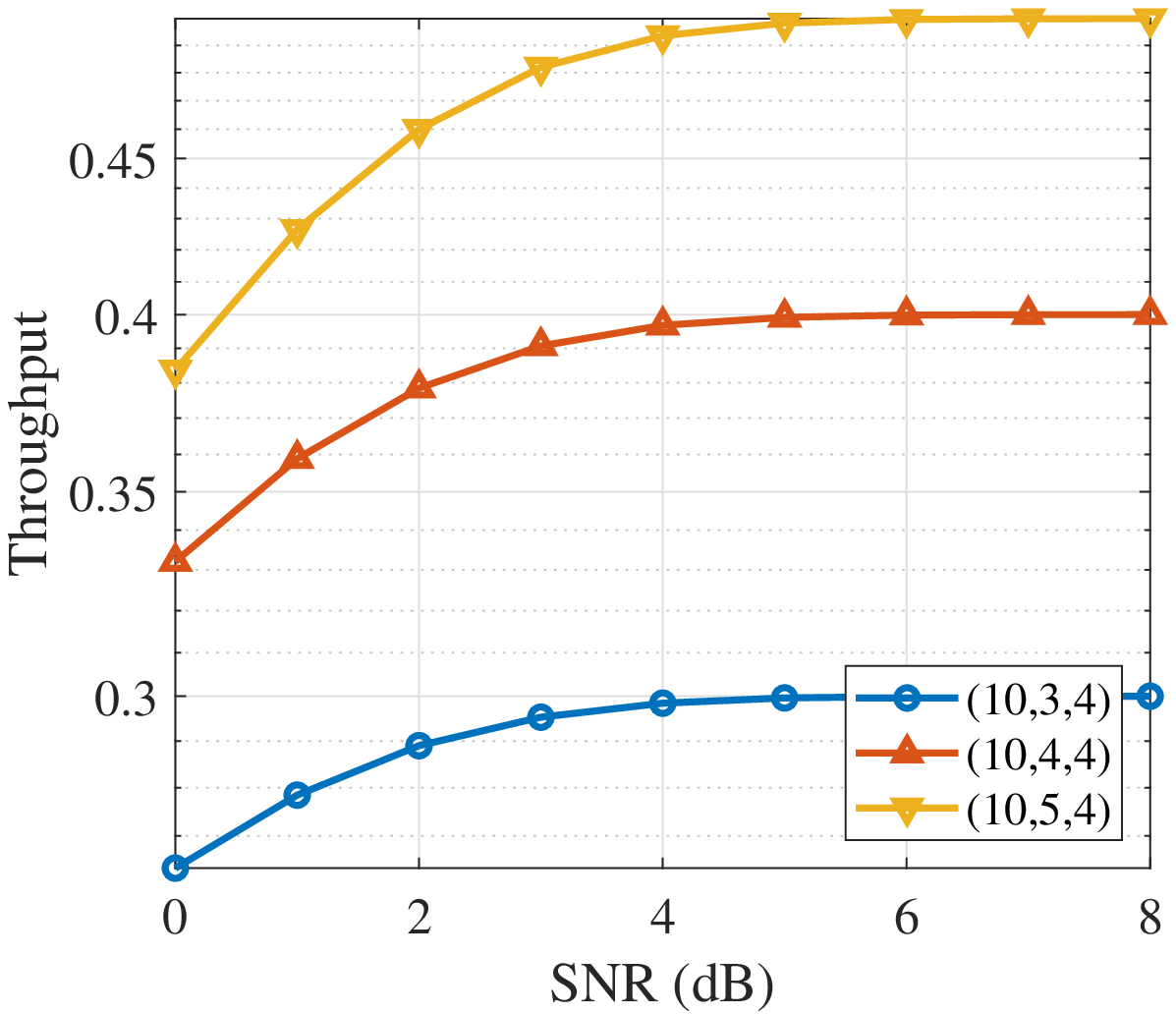}
\label{subfig:Tradeoff_k_throughput}\
}
\hspace*{-20pt}
\subfloat[]{
\includegraphics[width=0.55\linewidth] {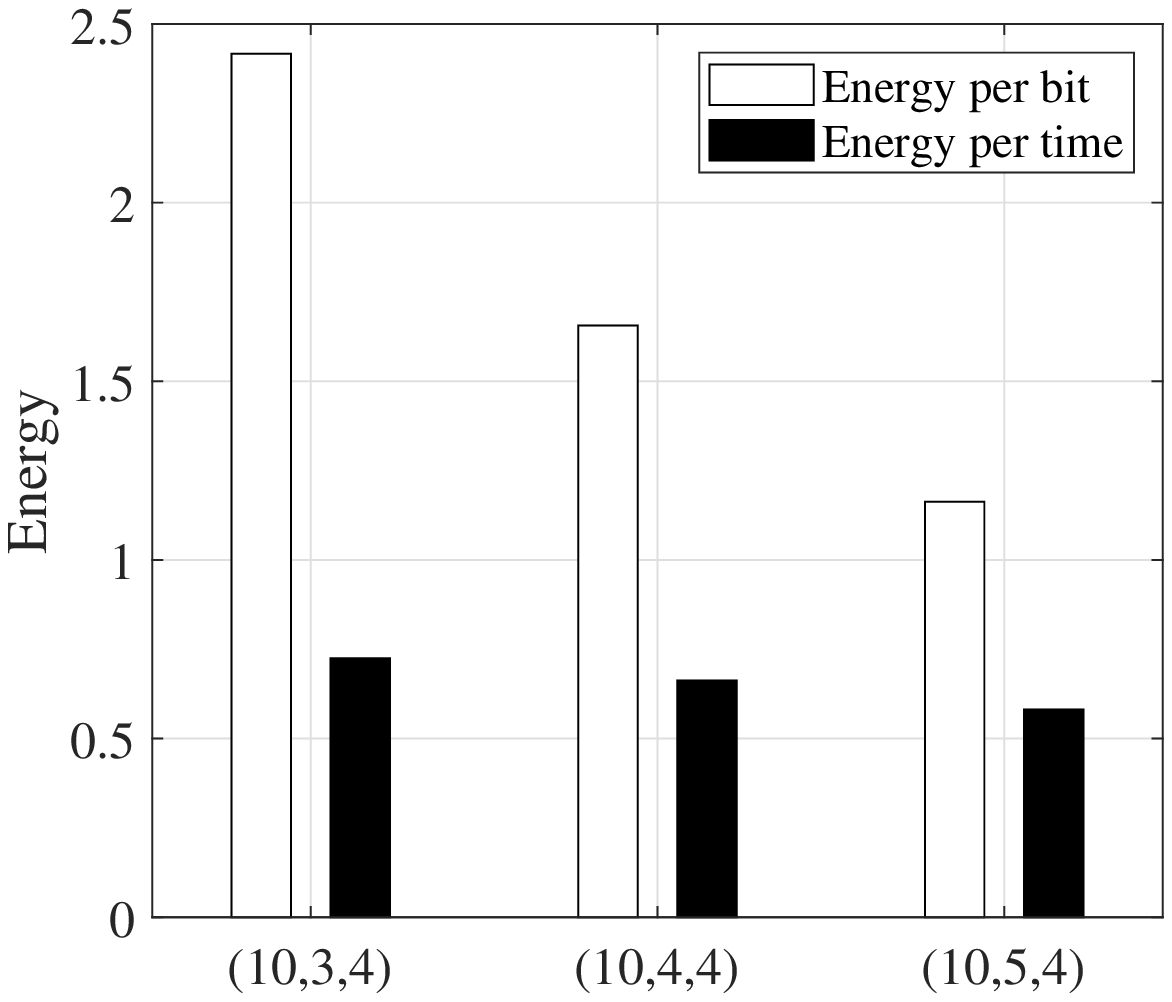}
\label{subfig:Tradeoff_k_energy}\
}
\hspace*{-70pt}
\caption{
Performance trade-off with the difference in the number of codewords $k$. (a) Throughput ${\eta _{10,k,4}}$. (b)  Energy per time, $Q_{10,k,4}^t$, and energy per bit, $Q_{10,k,4}^d$.
}
\label{fig:Tradeoff_k}
\end{figure}

In Fig. \ref{subfig:Tradeoff_k_throughput} and Fig. \ref{subfig:Tradeoff_k_energy}, the $(10,3,4)$, $(10,4,4)$, and $(10,5,4)$ codebooks are evaluated in terms of throughput and harvested energy, respectively, for different numbers of codewords $k$ with fixed values of $n$ and $d$.
The $(10,3,4)$ codebook yields the highest performance in terms of harvested energy and lowest throughput performance.
In contrast, the $(10,5,4)$ codebook yields the lowest performance in terms of harvested energy and highest performance in terms of throughput.
As $k$ increases, the throughput increases because the code rate increases while the harvested energy decreases; this is because the proposed codebook from the high-density codebook design algorithm comprises codewords that have a relatively fewer number of 1's to satisfy the minimum Hamming distance condition.
The harvested energy per time increases by a maximum of $100\%$, and the harvested energy per bit increases by a maximum of $25\%$, while the throughput decreases by a maximum of $40\%$.

\begin{figure}[!t]
\centering
\hspace*{-60pt}
\subfloat[]{
\includegraphics[width=0.55\linewidth] {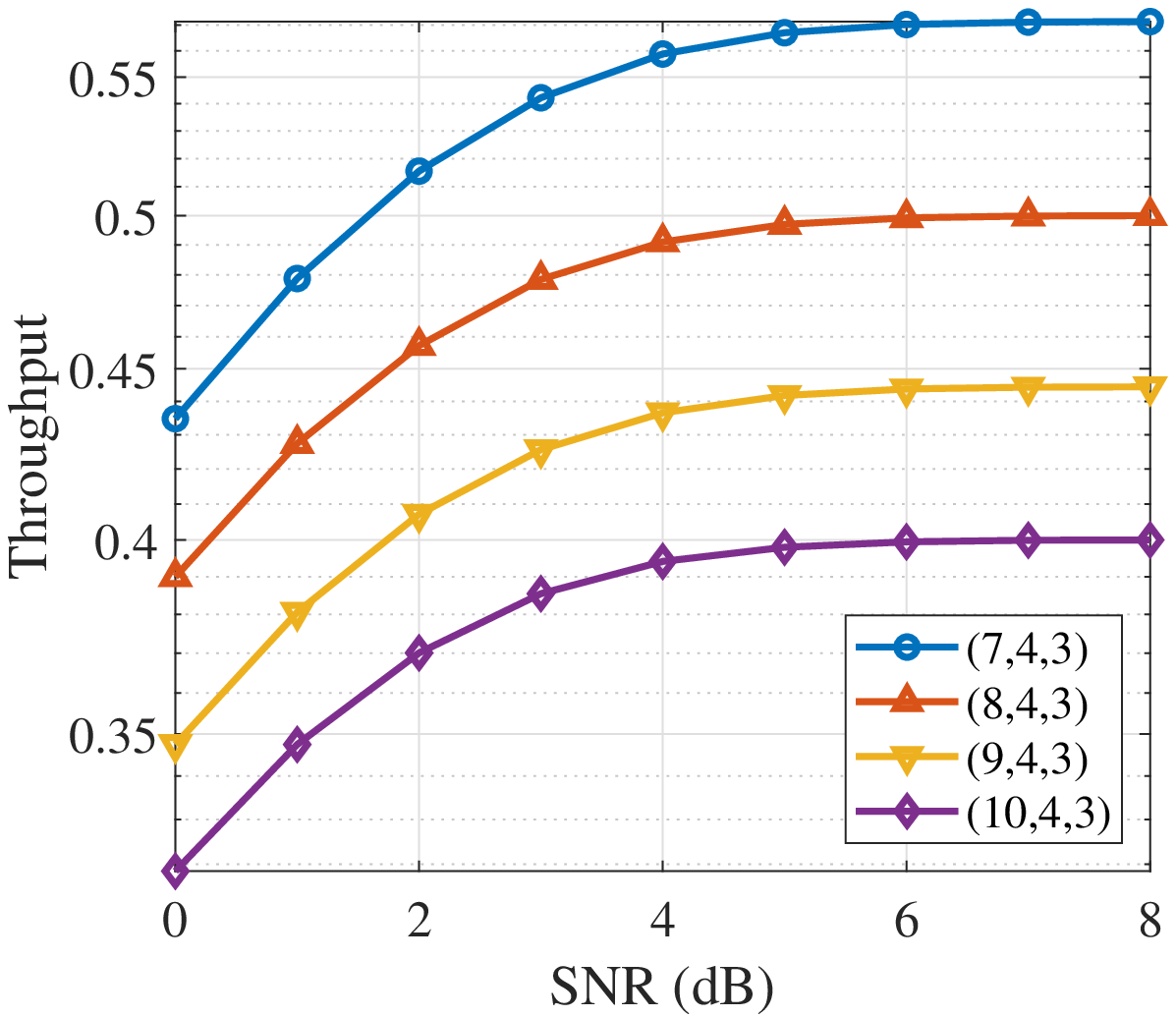}
\label{subfig:Tradeoff_n_throughput}\
}
\hspace*{-20pt}
\subfloat[]{
\includegraphics[width=0.55\linewidth] {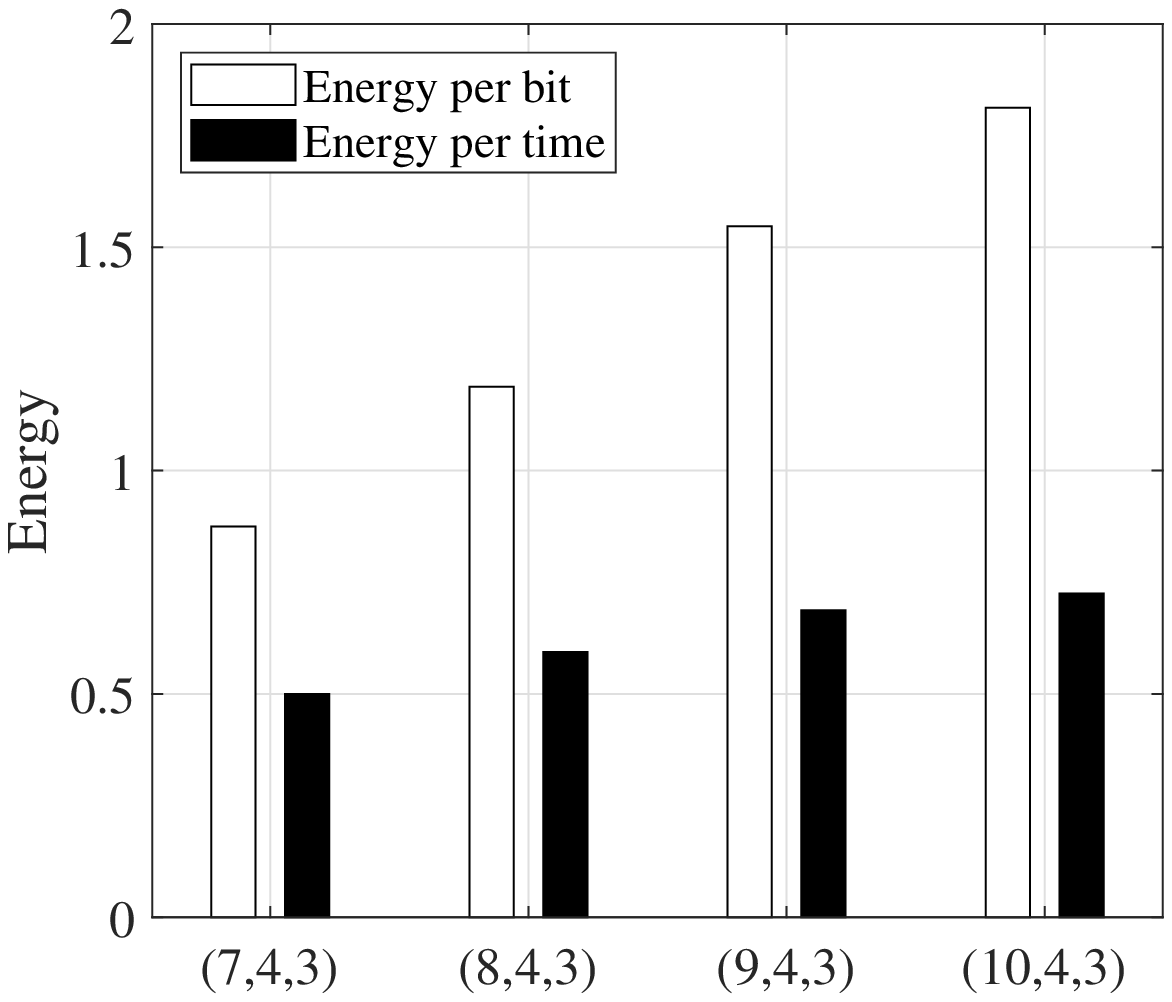}
\label{subfig:Tradeoff_n_energy}\
}
\hspace*{-70pt}
\caption{Performance trade-off with the difference in the length of codewords, $n$. (a) Throughput ${\eta _{n,4,3}}$. (b) Energy per time, $Q_{n,4,3}^t$, and energy per bit, $Q_{n,4,3}^d$.
}
\label{fig:Tradeoff_n}
\end{figure}

In Fig. \ref{subfig:Tradeoff_n_throughput} and Fig. \ref{subfig:Tradeoff_n_energy}, the $(7,4,3)$, $(8,4,3)$, $(9,4,3)$, and $(10,4,3)$ codebooks are evaluated in terms of throughput and harvested energy, respectively, for different codeword lengths $n$ with fixed values of $k$ and $d$.
The $(10,4,3)$ codebook yields the highest performance in terms of harvested energy and lowest performance in terms of throughput.
In contrast, the $(7,4,3)$ codebook yields the lowest performance in terms of harvested energy and highest performance in terms of throughput.
As $n$ increases, the throughput decreases because the code rate decreases while the harvested energy increases; this is because the number of 1's in the proposed high-density codebook increases as $n$ increases.
The harvested energy per time increases by a maximum of $110\%$, and the harvested energy per bit increases by a maximum of $40\%$, while the throughput decreases by a maximum of $30\%$.


\section{Conclusions}
\label{section:conclusions}

In this study, we proposed a high-density coding scheme that can control the throughput and harvested energy in a given channel using a genetic local search-based high-density codebook design algorithm that generates a high-density codebook.
We designed codebooks that satisfy the given $\left( {n,k,d} \right)$ condition, which is in general an NP-hard problem, and analyzed the throughput and harvested energy performances according to the variation in the $\left( {n,k,d} \right)$ condition in the theoretical results.
The simulation results revealed a trade-off between throughput and harvested energy. In addition, the throughput increased and harvested energy decreased as $k$ increased, whereas the throughput decreased and harvested energy increased as $n$ and $d$ increased.

\appendix
\numberwithin{equation}{section}
\section{Proof of BLER}
\subsection{Proof of equation~\ref{ref:BLER}}
The pairwise error probability that the decoder decides for a different codeword ${\mathbf{c}}_j$ when ${\mathbf{c}}_i$ is transmitted is derived as follows:
\begin{equation}
\begin{split}
 &P\left( {{\mathbf{c}}_i  \to {\mathbf{c}}_j } \right) \\
  &= P\left( {\left\| {{\mathbf{y}} - {\mathbf{c}}_j } \right\|^2  \le \left\| {{\mathbf{y}} - {\mathbf{c}}_i } \right\|^2 } \right), \\
  &= P\left( {\left\| {{\mathbf{n}} + \left( {{\mathbf{c}}_i  - {\mathbf{c}}_j } \right)} \right\|^2  \le \left\| {\mathbf{n}} \right\|^2 } \right). \\
  &= P\left( {\sum\limits_{r = 0}^{n - 1} {\left[ {n_r^2  + 2n_r \left( {c_{i,r}  - c_{j,r} } \right) + \left( {c_{i,r}  - c_{j,r} } \right)^2 } \right]}  \le \sum\limits_{r = 0}^{n - 1} {n_r^2 } } \right), \\
  &= P\left( {\sum\limits_{r = 0}^{n - 1} {n_r \left( {c_{i,r}  - c_{j,r} } \right)}  \le  - \frac{1}{2}\sum\limits_{r = 0}^{n - 1} {\left( {c_{i,r}  - c_{j,r} } \right)^2 } } \right) \\
  &= P\left( {\sum\limits_{r = 0}^{n - 1} {n_r \left( {c_{i,r}  - c_{j,r} } \right)}  \le  - d_H \left( {{\mathbf{c}}_i ,{\mathbf{c}}_j } \right) \cdot E_b } \right).
\end{split} \nonumber
\end{equation}
where ${\mathbf{y}}$ and ${\mathbf{n}}$ denote the received signal sequence and noise sequence, respectively, when ${\mathbf{c}}_i$ is transmitted.
${d_H \left( {{\mathbf{c}}_i ,{\mathbf{c}}_j } \right)}$ denotes the Hamming distance between the codewords ${\mathbf{c}}_i$ and ${\mathbf{c}}_j$.
The left-hand side of the inequality in the probability function is the sum of Gaussian random variables with zero mean and variance, ${N_0  \cdot E_b  \cdot d_H \left( {{\mathbf{c}}_i ,{\mathbf{c}}_j } \right)}$.
\begin{equation}
\begin{split}
 &P\left( {\sum\limits_{r = 0}^{n - 1} {n_r \left( {c_{i,r}  - c_{j,r} } \right)}  \le  - d_H \left( {{\mathbf{c}}_i ,{\mathbf{c}}_j } \right) \cdot E_b } \right), \\
 & = P\left( {\frac{{\sum\nolimits_{r = 0}^{n - 1} {n_r \left( {c_{i,r}  - c_{j,r} } \right)} }}{{\sqrt {N_0  \cdot E_b  \cdot d_H \left( {{\mathbf{c}}_i ,{\mathbf{c}}_j } \right)} }} \ge \frac{{d_H \left( {{\mathbf{c}}_i ,{\mathbf{c}}_j } \right) \cdot E_b }}{{\sqrt {N_0  \cdot E_b  \cdot d_H \left( {{\mathbf{c}}_i ,{\mathbf{c}}_j } \right)} }}} \right) \\
 & = Q\left( {\frac{{d_H \left( {{\mathbf{c}}_i ,{\mathbf{c}}_j } \right) \cdot E_b }}{{\sqrt {N_0  \cdot E_b  \cdot d_H \left( {{\mathbf{c}}_i ,{\mathbf{c}}_j } \right)} }}} \right), \\
 & = Q\left( {\sqrt {d_H \left( {{\mathbf{c}}_i ,{\mathbf{c}}_j } \right) \cdot \frac{{E_b }}{{N_0 }}} } \right).
\end{split} \nonumber
\end{equation}

The codewords with the minimum Hamming distance to the transmitted codeword dominate the codeword error probability.
Therefore, the codeword error probability is approximated as below.

\begin{equation}
\begin{split}
 P_{n,M,d}  &\approx \sum\limits_{i = 0}^{2^k  - 1} {p\left( {{\mathbf{c}}_i } \right)A_{d,i}  \cdot Q\left( {\sqrt {d_H \left( {{\mathbf{c}}_i ,{\mathbf{c}}_j } \right) \cdot \frac{{E_b }}{{N_0 }}} } \right)}  \\
 & \approx \sum\limits_{i = 0}^{2^k  - 1} {\frac{1}{{2^k }} \cdot A_{d,i}  \cdot Q\left( {\sqrt {d \cdot \frac{{E_b }}{{N_0 }}} } \right)}  \\
 \end{split} \nonumber
\end{equation}

\end{document}